# Non-trivial Bifocal and Optical Vortex Generation of DNG Materials Unveiled by Generalized Transfer Matrix Method and Matrix Fourier Optics

Yifeng Qin

*Peng Cheng Laboratory, Shenzhen, Guangdong, 518052, China*

The investigation and analysis of exotic physical phenomena facilitated by metamaterials have emerged as a compelling area of interest in physics, material science, and engineering. However, there remains a lack of suitable theoretical tools for scrutinizing light-matter interactions in metamedia, particularly those involving distinct constitutive tensors, such as polarization cross-coupling or non-reciprocal propagation induced by bianisotropy. In this research, we apply the generalized transfer matrix method and matrix Fourier Optics to devise a novel hybrid methodology adept at efficiently tracing an optical beam as it propagates through transverse-homogeneous, longitudinal-inhomogeneous bianisotropic media. Our proof-of-concept demonstration elucidates the interaction between a nonparaxial Gaussian incident beam with various linear/circular polarizations and a mismatched anisotropic double-negative (DNG) metamaterial. We unveil the nontrivial bifocal effect and optical vortex (OV) generation and focusing phenomena in a three-dimensional environment, findings that have not been documented previously. Significantly, the proposed efficient methodology holds the potential for application in examining more promising physical phenomena exhibited by intricate metamaterials, enabling a more visually rigorous approach.

**Keywords:** Bifocal; Optical vortex generation; Bianisotropy; Visualized hybrid methodology.

## I. INTRODUCTION

Over recent decades, metamaterials and metasurfaces have garnered substantial attention due to their distinctive properties, such as perfect lensing, anomalous reflection, beam manipulation, and polarization filtering [1-6]. These exceptional phenomena offer considerable potential for industrial applications, encompassing spectrometers, next-generation communications, and virtual reality (VR)/augmented reality (AR) systems [7-12]. The primary emphasis of metamaterial research has been the development of functional metadevices through a two-step process: initially engineering the subunits (meta-atoms) and subsequently assembling them in an inhomogeneous manner based on a predefined phase profile, assuming normal plane wave incidence. Nonetheless, the response of meta-atoms inherently depends on various factors, including wavelength, polarization, spin, and angle of the incident light. The responses of homogeneous metamaterials to diverse polarized optical beams containing non-scalar momentum components can also yield fascinating physical phenomena, such as optical vortex (OV) beam generation [13-18]. As a result, the underlying interactions between optical beams and metamaterials remain an enticing subject for further scholarly inquiry.

The design of meta-atoms, crucial for metamaterial functionalities, largely depends on prior knowledge. For example, it is known that metallic wires exhibit negative permittivity below the plasmon frequency [19], while induced resonators (e.g., split-ring resonators (SRRs)) possess *bianisotropy* due to magnetoelectric coupling [20]. Identifying the properties of meta-atoms with constitutive tensors enables researchers to study complex physical phenomena without considering specific structural realizations. Furthermore, understanding the connection between the underlying mechanisms of observed complex phenomena and the constitutive parameters of metamaterials can aid in designing functional meta-atoms. The effective medium theory (EMT) is a homogenization method that represents meta-atoms as constituents with effective permittivity and permeability tensors [21-38], efficiently evaluating the macroscopic properties of multilayer dielectric bulk materials. However, EMT is inapplicable to meta-surfaces exhibiting "phase jumps," such as surface plasma polaritons, due to irrationally large nominal permittivity and permeability values. Moreover, certain critical parameter regimes mixing evanescent and propagating light transport may counterintuitively amplify nonlocal effects, causing the EMT approximation to break down. An alternative approach for efficiently modeling metasurfaces is the equivalence surface theorem, which homogenizes electric and magnetic surface polarization densities as generalized sheet transition conditions (GSTCs) [4, 39-45]. Nonetheless, this model neglects metasurfaces' longitudinal effect, potentially leading to significant inaccuracies at oblique incidence.

In this study, we present the generalized transfer matrix method (GTMM), a sophisticated approach capable of analyzing the response of multilayered *bianisotropic* media to incident waves with diverse polarizations at arbitrary incidence angles. Conventional EMT-based TMM predominantly addresses a material's macroscopic behaviors, with the majority of prior research incorporating bianisotropic parameters into permittivity and permeability to ascertain the ultimate transmittance and reflectance [28, 35, 37]. In contrast, our methodology subdivides bulk materials into numerous slender layers, facilitating the examination of both macroscopic and microscopic properties of metamaterials (refer to FIG.1). Notably, the gyration pseudotensors $\bar{\bar{\xi}}$ and $\bar{\bar{\zeta}}$, which delineate the mutual coupling between electric and magnetic fields, are integrated into our formulas to scrutinize the material's bianisotropy and to facilitate a thin metafilm's exhibition of a pronounced phase change from the incident surface to the exit surface. The eigenmodes of each metamaterial stratum are rigorously sorted and compelled to comply with momentum and electromagnetic field boundary conditions, ensuring the method's stability and granting access to reflection and transmission coefficient Jones matrices rather than transmittance and reflectance. The eigen-states of momentum components in the incident region determine the coordinate system under consideration and serve a critical function as normalization operators. Matrix Fourier Optics is employed to address the spatially-defined nonparaxial source beam with various states of polarizations (SOPs), converting the source from the spatial domain to the momentum domain while preserving polarization information. The integration of the GTMM and matrix Fourier Optics allows for the efficient tracing of the optical beam's propagation while retaining its full wave nature.

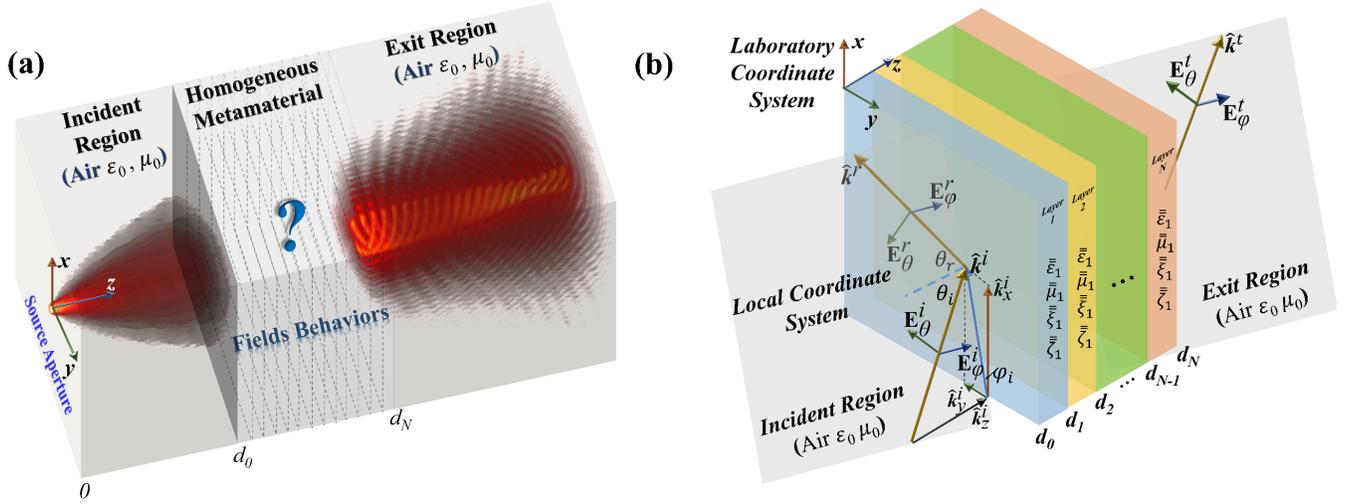

FIG.1. (a) The general conceptual model that our methodology targets. (b) The geometry of the considered multi-layer model for GTMM. The superscripts *i*, *r*, and *t* stand for incident, reflect, and transmit, respectively.

This paper is structured as follows. In Section II, we present the theoretical framework of our hybrid methodology and showcase its remarkable computational efficiency. We address salient challenges in resolving the system depicted in FIG.1, which include accurately defining the light beam's SOP, attaining efficient outcomes for diverse coordinate systems, and ascertaining boundary conditions. In Section III, the methodology is utilized to investigate the interaction between a nonparaxial Gaussian beam with different SOPs and a generic double-negative index (DNG) material. Remarkably, a non-trivial bifocal effect is discerned both inside and outside the DNG material when the incident optical beam is vertically (*y*-) polarized. Additionally, we scrutinize the spin-orbit interaction (SOI) between a circularly polarized (CP) Gaussian beam and the DNG material. The anticipated OV generation emerges for the cross-polarized beam, and most importantly, the simulation demonstrates that the rotational orientations of the generated OV beam's wavefront exhibit opposite directions in the DNG material and free space. In conclusion, we consolidate our key contributions and offer an overview of the potential applicability of the proposed methodology in the investigation of other intriguing physical phenomena associated with metamaterials.

## II. OVERVIEW OF THE METHODOLOGY

### A. Generalized TMM

The general bianisotropic material takes a constitutive relation as follows:

$$\mathbf{D} = \bar{\bar{\varepsilon}} \mathbf{E} + \bar{\bar{\xi}} \mathbf{H}, \quad (1.1)$$

$$\mathbf{B} = \bar{\bar{\zeta}} \mathbf{E} + \bar{\bar{\mu}} \mathbf{H}, \quad (1.2)$$

where **D**, **B**, and **H** are the electric-displacement vector, magnetic-displacement vector, and magnetic fields, respectively. Additionally, $\bar{\bar{\varepsilon}}$ and $\bar{\bar{\mu}}$ are permittivity and permeability tensors, while $\bar{\bar{\xi}}$ and $\bar{\bar{\zeta}}$ are gyration pseudo-tensors.

In our consideration and most experimental setups, the normal vector of the multilayer planar meta-slabs or samples under test aligns with the *z*-axis. Consequently, we present all formulas and derivations in the laboratory coordinate system first, which facilitates handling the boundary conditions at each interface. Subsequently, we introduce an efficient method to convert the Jones matrices between laboratory and local coordinates.

The state equations for transverse electric and magnetic fields with a certain surface vector $\hat{\mathbf{k}}_s$ are as follows:

$$\frac{d}{dz}\mathbf{V}(z) = \bar{\bar{\mathbf{M}}}\mathbf{V}(z), \quad (2)$$

where $\mathbf{V} = [E_x, E_y, H_x, H_y]^T$, and the elements of the state matrix $\bar{\bar{\mathbf{M}}}$ are dependent on $\bar{\bar{\varepsilon}}$, $\bar{\bar{\mu}}$, $\bar{\bar{\xi}}$ and $\bar{\bar{\zeta}}$. Although other scholars have presented the expression of $\bar{\bar{\mathbf{M}}}$ [37], we re-derived this expression in an alternative way to obtain the fundamental solution in terms of the eigenmatrix and eigenvalues, which follows:

$$\mathbf{V}(z) = \bar{\bar{a}}\exp(\bar{\bar{\lambda}}z)\mathbf{A}, \quad (3)$$

where $\bar{\bar{a}}$ and $\lambda_i$s are the eigenmatrix and eigenvalues of $\bar{\bar{\mathbf{M}}}$, respectively. The vector $\mathbf{A}$ is the weight vector. From a physical standpoint, the eigenvalues $\lambda_i$s are the material's longitudinal wave vector $k_z$s for the input surface wave vector ($k_x$, $k_y$) (phase match condition). Thus, by carefully *sorting the modes*, the breakdown of the GTMM can be avoided.

The transfer relationship between two different planes can be expressed as follows:

$$\mathbf{V}(z) = \bar{\bar{P}}(z, z')\mathbf{V}(z'), \quad (4)$$

where $\bar{\bar{P}}(z, z') = \bar{\bar{a}}\exp(\bar{\bar{\lambda}}(z-z'))\bar{\bar{a}}^{-1}$. Applying the electro-magnetic (EM) tangential fields boundary conditions, we can have the general transfer formula from the incident interface ($z = d_0$) to the exit interface ($z = d_N$):

$$\bar{\bar{a}}_c^{-1}|_{\text{exit}}\prod_{i=1}^{N}\bar{\bar{P}}(d_i, d_{i-1})\bar{\bar{a}}_c|_{\text{inci}}\begin{bmatrix}\bar{\bar{R}}\\ \bar{\bar{I}}\end{bmatrix} = \begin{bmatrix}\bar{\bar{0}}\\ \bar{\bar{T}}\end{bmatrix} \quad (5)$$

where $\bar{\bar{I}}$ and $\bar{\bar{0}}$ are the 2 by 2 identity and zero matrix, respectively. $\bar{\bar{a}}_c$ is the eigenmatrix of the incident and exit region under the cartesian coordinate system, which is can be directly achieved if the region is air. Based on above equations, the reflection and transmission Jones matrices $\bar{\bar{R}}$ and $\bar{\bar{T}}$ can be derived. The detailed derivation processes are shown in Supplemental Material [46].

The equations shown above can also be utilized to calculate the reflection and transmission Jones matrices for the TE and TM waves by only replacing the eigenstate matrix $\bar{\bar{a}}_c$ with a new eigenstate matrix $\bar{\bar{a}}_m$. The $\bar{\bar{a}}_m$ is incident/exit region's eigenstate matrix considered in local coordinate system or momentum space. The $\bar{\bar{a}}_m$ and $\bar{\bar{a}}_c$ can be straightforwardly transformed to each other using the conversion relation:

$$\begin{bmatrix}\hat{\mathbf{e}}_x\\ \hat{\mathbf{e}}_y\end{bmatrix} = \begin{bmatrix}\cos\theta\cos\varphi & \sin\varphi\\ \cos\theta\sin\varphi & -\cos\varphi\end{bmatrix}\begin{bmatrix}\hat{\mathbf{e}}_\parallel\\ \hat{\mathbf{e}}_\perp\end{bmatrix}. \quad (6)$$

Essentially, the eigenstate matrix of the incident/exit region serves as a normalization operator, and the transfer matrices signify the eigenstates of waves in materials that satisfy the boundary conditions.

If we specify $\bar{\bar{R}} = \begin{bmatrix}R_{uu} & R_{uv}\\ R_{vu} & R_{vv}\end{bmatrix}$ and $\bar{\bar{T}} = \begin{bmatrix}T_{uu} & T_{uv}\\ T_{vu} & T_{vv}\end{bmatrix}$, where subscript $u$ represents $x$ or $\parallel$, and $v$ represents $y$ or $\perp$, the formula of Jones matrices for CP waves in Cartesian space or momentum space can be expressed as:

$$\bar{\bar{R}}_{CP} = \frac{1}{2}(R_{uu} + R_{vv})\bar{\bar{I}} + \frac{i}{2}(R_{uv} - R_{vu})\bar{\bar{\sigma}}_3$$

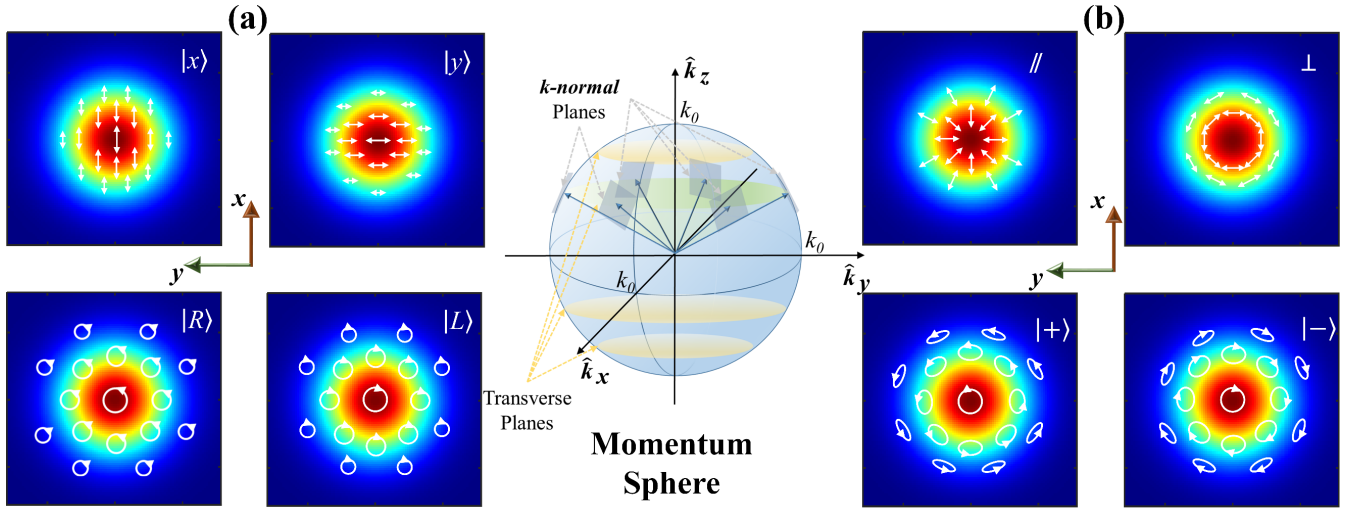

FIG. 2. Eight different polarized source Gaussian beams (a) The *x*-, *y*-, right-hand circularly and left-hand circularly polarized (RCP and LCP) Gaussian beams in the **spatial** domain. (b) The transformed **spatial** TM ($\parallel$), TE ($\perp$), RCP (+) and LCP (-) Gaussian beams that are originally defined in the momentum space. The central sphere, defined in the momentum domain, encompasses the components that can propagate in free space. The polarizations of fields specified in the momentum domain lie within *k*-normal planes, whereas the polarizations of spatially-defined fields are situated in transverse planes.

$$+\frac{1}{2}(R_{uu}+R_{vv})\,\bar{\bar{\sigma}}_1 + \frac{1}{2}(R_{uv}+R_{vu})\,\bar{\bar{\sigma}}_2, \qquad (7.1)$$

$$\bar{\bar{T}}_{CP} = \frac{1}{2}(T_{uu}+T_{vv})\,\bar{\bar{I}} - \frac{i}{2}(T_{uv}-T_{vu})\,\bar{\bar{\sigma}}_3$$

$$+\frac{1}{2}(T_{uu}-T_{vv})\,\bar{\bar{\sigma}}_1 - \frac{1}{2}(T_{uv}+T_{vu})\,\bar{\bar{\sigma}}_2. \qquad (7.2)$$

Now, all the necessary information for computing the interaction of optical fields with different SOPs and a metamaterial is accessible. The GTMM is verified and examples are shown in Supplemental Material [46].

### B.  Polarized Gaussian Beam and Matrix Fourier Optics

Ray optics [47-48] is the most popular analytical method to investigate the wave behaviors through various meta-materials. However, as an approximate high-frequency method, it loses efficiency when the target model requires iterative processing and fails to account for the complex physics offered by meta-atoms. In contrast, Fourier Optics decomposes a spatial optical source into plane waves propagating in different directions, transforming a problem from spatial space to momentum space. By handling all plane wave components collectively, Fourier Optics captures sufficient full-wave physics. The well-known Fourier optics theorem can be summarized using the following two equations [49]:

$$U(k_x, k_y) = \iint E(x,y)\,e^{i(k_x x + k_y y)}\mathrm{d}x\mathrm{d}y, \qquad (8.1)$$

$$E(x, y) = \frac{1}{4\pi^2}\iint U(k_x,k_y)\,e^{-i(k_x x + k_y y)}\mathrm{d}k_x\mathrm{d}k_y \qquad (8.2)$$

where $E(x, y)$ is the field-distribution in the spatial domain, and $U(k_x, k_y)$ represents the weight coefficients of plane waves propagating along the wave vector $(k_x, k_y)$.

However, the traditional Fourier optics transform is a scalar process, and its underlying rule is that the transformed function $U(k_x, k_y)$ only represents the components lying in the transverse plane in the momentum space. Also, the SOP of $E$ and $U$ are identical but in spatial and momentum space, respectively. Therefore, more comprehensive trans-formation formulas are necessary that takes into account the SOP information. In this paper, without loss of generality, we employ a Gaussian beam as the incident source, while, similar to the method proposed in [50], we define the Gaussian *incident* beam in a matrix form. Defined in spatial domain, the source Gaussian beam with different SOPs which is consistent with laboratory coordinate system $(x, y)$, are expressed as

$$\mathbf{E}^{inci}_{c|s}(x, y, 0) = e^{-\frac{(x^2+y^2)}{w_0^2}}\,\bar{\bar{\psi}}_s\begin{bmatrix}\hat{\mathbf{e}}_\mathbf{x}\\ \hat{\mathbf{e}}_\mathbf{y}\end{bmatrix}, \qquad (9.1)$$

$$\mathbf{U}^{inci}_{c|s}(k_x, k_y, 0) = \iint \mathbf{E}^{inci}_{c|s}(x, y, 0)\,e^{i(k_x x + k_y y)}\mathrm{d}x\mathrm{d}y, \qquad (9.2)$$

where $\bar{\bar{\psi}}_s$ are the eigenmatrices of the Pauli matrices $\bar{\bar{\sigma}}_s$, and $s = 1, 2,$ and 3. When $s = 1$, the SOPs correspond to the Stokes parameter $p_1$, which are linearly polarized (LP) at 0° and 90°. Similarly, when $s = 3$, the polarizations are CP. To simplify our analysis, the 45° and 135° LP states represented by the Stokes parameter $p_2$ are excluded in our discussion. The subscript $c$ indicates that the incident optical beam is originally defined in the Cartesian coordinate system, and $w_0$ is the beam-waist. The total four polarization states of the Gaussian source beam in spatial space are illustrated in FIG.2(a). In addition, the transverse wavenumbers $k_x$ and $k_y$ are located in the range $[-k_0, k_0]$ and $\sqrt{k_x^2 + k_y^2} \leq k_0^2$.

For an oblique incident plane wave, the transverse electric (TE, ⊥) and the transverse magnetic (TM, ∥) polarizations are the two eigen-SOPs that can physically describe the behaviors of the waves at an interface, that is, reflection and transmission. Therefore, it is more straightforward to define these two states in the local coordinate system, as shown in FIG.1. Specifically, $E_\theta$ is a TM wave and $E_\varphi$ represents a TE wave. Thus, we can alternatively define the Gaussian source beam directly in the momentum space and obtain the spatial distributions with a Fourier transform. The expressions are as follows:

$$\mathbf{U}^{inci}_{m|s}(k_x, k_y, 0) = e^{-\frac{(k_x^2+k_y^2)w_0^2}{4}}\,\bar{\bar{\psi}}_s\begin{bmatrix}\hat{\mathbf{e}}_\parallel\\ \hat{\mathbf{e}}_\perp\end{bmatrix}, \qquad (10.1)$$

$$\mathbf{E}^{inci}_{m|s}(x, y, 0) = \frac{w_0^2}{4\pi}\iint \mathbf{U}^{inci}_{m|s}(k_x, k_y, 0)\,e^{-i(k_x x + k_y y)}\mathrm{d}k_x\mathrm{d}k_y, \qquad (10.2)$$

where the subscript $m$ indicates that the incident optical beam is originally defined in the momentum space. The four states of the Gaussian source beam in spatial space are illustrated in FIG.2(b). In addition, the ideal CP waves defined in momentum space are elliptically polarized waves observed in a spatial-transverse plane because of the projection from $\boldsymbol{k}$-normal planes to the spatial-transverse plane, and vice versa. This phenomenon leads to an interesting SOI effect, which is discussed in Section III.

Upon properly defining the source beam, the spatial field distribution on the plane $z = z'$ in free space can be predicted by multiplying a phase component $e^{-ik_z z'}$ by the momentum component $\mathbf{U}(k_x, k_y)$ and subsequently performing an inverse Fourier transform.

Moreover, the *reflected* and *transmitted* waves by the multilayer metamaterials in free space can be determined using the following expressions:

$$\mathbf{E}^{refl}_{m,c|s}(x, y, z') = \iint \bar{\bar{R}}_{m,\,c|s}(k_x, k_y, d_0)\,\mathbf{U}^{inci}_{m,c|s}(x, y, z')$$
$$\cdot e^{-i(k_x x + k_y y) + ik_z z'}\mathrm{d}k_x\mathrm{d}k_y, \qquad (11.1)$$

$$\mathbf{E}^{tran}_{m,c|s}(x, y, z') = \iint \bar{\bar{T}}_{m,\,c|s}(k_x, k_y, d_0)\,\mathbf{U}^{inci}_{m,c|s}(k_x, k_y)$$
$$\cdot e^{-i(k_x x + k_y y) + ik_z z'}\mathrm{d}k_x\mathrm{d}k_y, \qquad (11.2)$$

where $\bar{\bar{R}}$ and $\bar{\bar{T}}$ are the reflection and transmission Jones matrices, respectively. A coefficient constant $w_0^2/4\pi$ is necessary if the source field is defined in momentum space.

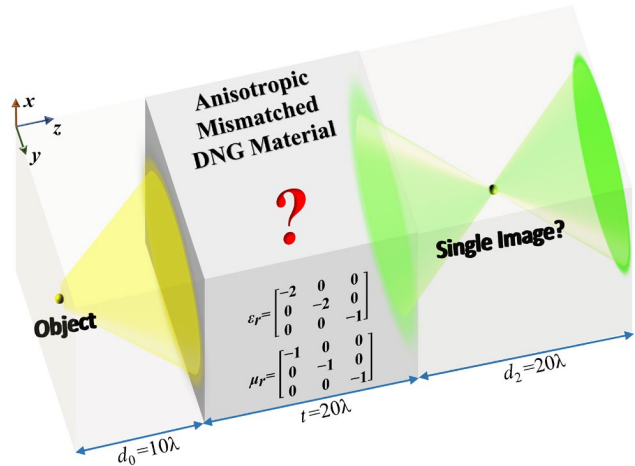

FIG.3.  Geometry of the considered general DNG model.

Both the Matrix Fourier optics and the GTMM employed in our hybrid method were analytical or semi-analytical. Consequently, our proposed hybrid method was highly computationally efficient compared to other numerical methods. Moreover, since the contribution of each plane wave component to the optical beam in the space of interest was adequately processed, the simulated results preserved their wave propagation nature. High-order momentum components beyond the range [-$k_0$, $k_0$] are filtered since we only focus on the beam's propagation properties.

## III.  METHODOLOY'S DEMONSTRATION

### A.  The Considered Targeting Model

The DNG metamaterial has been a prominent research topic due to its potential for exhibiting perfect lensing [1, 51-67]. Recently, novel concepts and techniques have been proposed for realizing DNG materials [68]. Due to the process constraints, no perfect 3D centrosymmetric DNG meta-atoms have been reported to date [69-70], indicating that all known DNG materials are anisotropic. However, to the best of our knowledge, no thorough investigation has been conducted on researching the interaction between a nonparaxial beam with diverse SOPs and an anisotropic DNG metamaterial. In this section, we apply the proposed methodology to examine a nonparaxial Gaussian beam's behaviors after it interacts with an *anisotropic*, *mismatched* DNG material, resulting some intriguing phenomena. Furthermore, we demonstrate the efficacy of the methodology in predicting these complex interactions.

Figure 4 illustrates the system under investigation. As this study aims to unveil the phenomena, all discussions pertain to a single frequency $f$, with the corresponding wavelength and wavenumber denoted as $\lambda$ and $k_0$, respectively. To simulate the point source, the beam waist of the incident Gaussian beam is set to be sub-wavelength, $w_0 = 0.5\lambda$. In order to better observe the focusing performance, the source aperture is placed $10\lambda$ away from the first interface, allowing the beam to sufficiently expand before entering the slab. The material slab has a thickness of $20\lambda$ to ensure the capture of all full-wave physics.

The permittivity and permeability tensors applied in our target model are depicted in FIG.3. It can be observed that the slab under consideration is a uniaxial anisotropic material, in which the transverse refractive index $n_s$ and the longitudinal refractive index $n_z$ differ. With the assumption of a normal incident plane wave, the material's state equations are significantly simplified, leading to the derivation of the characteristic impedances for the $x$-polarized and $y$-polarized waves, which are $\eta_0\sqrt{\varepsilon_{xx}/\mu_{yy}}$ and $\eta_0\sqrt{\varepsilon_{yy}/\mu_{xx}}$, respectively, where $\eta_0$ is the characteristic impedance of air. If $\varepsilon_{yy} = \mu_{xx}$ and $\varepsilon_{xx} = \mu_{yy}$, the impedance of the anisotropic material is transversely matched to air. Strictly speaking, this impedance matching condition is valid only for a normally incident plane wave. However, oblique incident components also exhibit high transmission efficiencies under the aforementioned condition.

At last, in [71], the authors presented propagation conditions for anisotropic negative index materials. We demonstrated that the DNG material in this study does not eliminate any momentum components of the beam (see Supplemental Material [46]).

### B.  Vertical ($y$-), TE and TM Polarized Gaussian Beam Interacts with DNG Metamaterial

First, we study the response of the DNG slab to a $y$-polarized Gaussian beam. At the source aperture, the distribution of the fields is $\mathbf{E}^{inci}(x, y, 0) = \exp(\frac{-x^2-y^2}{w_0^2})\hat{\mathbf{e}}_y$ based on Eq. (9). The field distributions predicted by our proposed hybrid method are illustrated in FIG.4. Specifically, Figure 4a presents the magnitude of the $y$-polarized electric field's distributions in the $x$-$z$ plane in spatial space, while FIG.4b shows the $|E_y|$ in 3D space.

We observe that the cone-shaped Gaussian beam focused twice inside the DNG bulk, with two corresponding focal beams in the exit region. Within the metamaterial, the full-width-at-half-maximum (FWHM) of the first focal beam was much narrower than the FWHM located on the right side of the middle plane ($z' = 20\lambda$), and the outgoing beam exhibited an opposite behavior in the exit region. This phenomenon seems counterintuitive at first glance, but it actually reflects the propagation of an LP optical beam for the Cartesian coordinate system in a 3D environment. Specifically, the $y$-polarized wave is an in-plane wave in the $y$-$z$ plane and a normal-plane wave in the $x$-$z$ plane. Moreover, the momentum components travel along arbitrary oblique directions, neither in the $x$-$z$ plane nor the $y$-$z$ plane, possess both in-plane and normal-plane components for the local coordinate system. In other words, the result shown in FIG.5 arises from a superposition of optical effects associated with both TE and TM waves.

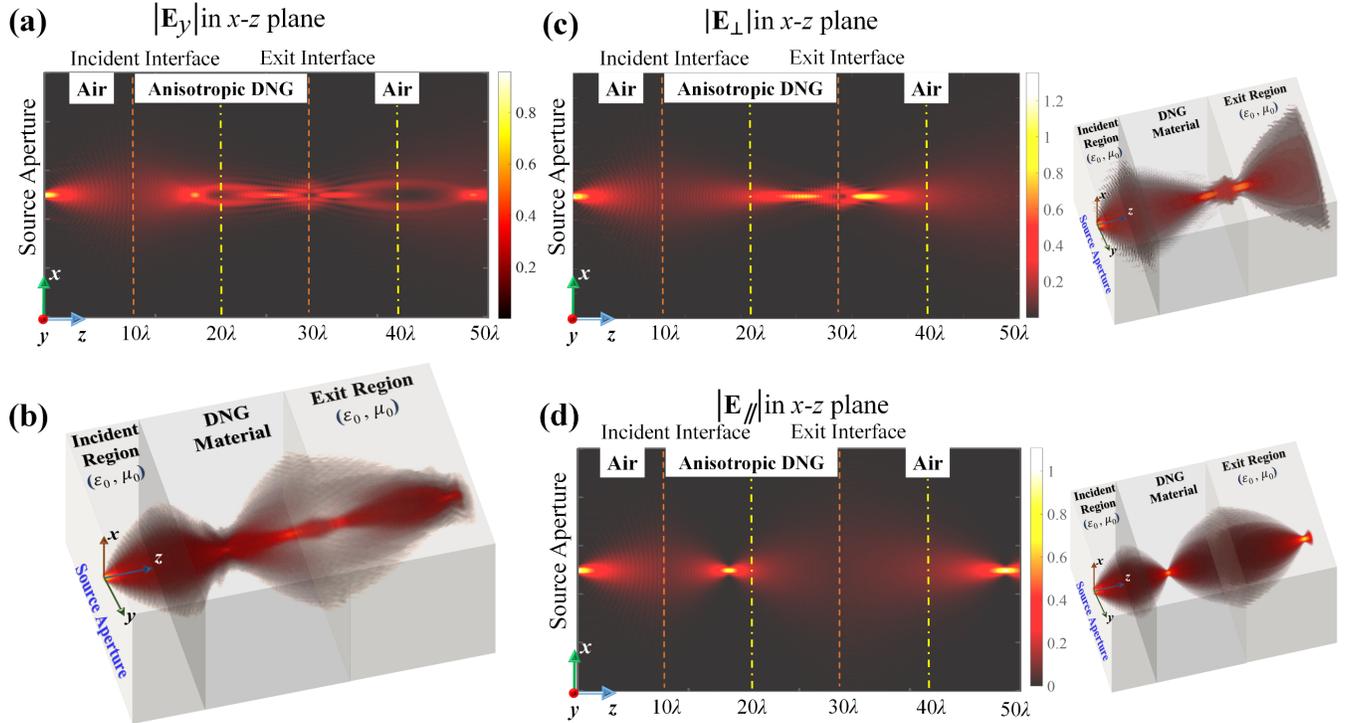

FIG.4. (a) The amplitude of *y*-polarized *E*-fields in the *x-z* plane across the whole system. (b) The 3D vision of the wave's propagation. (c) The TE-polarized wave's amplitude. (d) The TM-polarized wave's amplitude in the *x-z* plane across the whole system.

To verify the hypothesis, we trace the purely TE-wave and TM-wave propagations independently. We define the formulas of the source fields based on Eq. (10):

$$\mathbf{U}_{TM,TE}^{inci}(k_x, k_y, 0) = e^{\frac{-(k_x^2+k_y^2)w_0^2}{4}} \begin{bmatrix} \hat{\mathbf{e}}_{\parallel} \\ \hat{\mathbf{e}}_{\perp} \end{bmatrix}. \quad (12)$$

The simulation results are illustrated in FIG.4(c) and (d), aligning with our previous discussion. Each single SOP incident beam has only one focal point inside the DNG metamaterial and refocused once in the exit region. Inside the DNG bulk, the focal point attributed to the TM wave is located to the left of the metamaterial's central plane, while the focal point generated by the TE wave appears on the right. In addition to the rigorous mathematical proof, the divergence of focal points could be more intuitively interpreted. The TM wave possesses a *z*-component and is more sensitive to the transverse-index and longitudinal-index contrast. In these two test cases, the polarizations and field distributions are perfectly axis-rotating symmetric, rendering them essentially two 2D scenarios.

While previous studies have explored the unique behavior of TE and TM waves when interacting with anisotropic negative refractive index materials [52, 54, 64], we present, to the best of our knowledge, the first report on the bifocal physics of DNG metamaterials. To generate and observe this phenomenon, several crucial prerequisites must be met: the DNG material must exhibit an appropriate refractive index contrast between the transverse and longitudinal refractive indices; the meta-slab thickness must be adequately large; the impedances cannot be perfectly matched; and the observation range must be sufficiently extensive. Without an effective analytical tool, certain physical phenomena might remain hidden from observation. This section effectively demonstrates that the proposed methodology serves as a valuable resource for researchers in the field, facilitating a deeper understanding of metamaterials.

### C. OV Beam's Generation and Focusing

We have successfully demonstrated the general DNG metamaterial's response to an LP Gaussian beam. This naturally leads to the question of how a CP Gaussian beam would behave in the considered system. Since a CP wave can be viewed as a specific combination of TE and TM components, we anticipated the bifocal effect for the co-polarized beam. However, predicting the propagation behavior of the cross-polarized beam, which arises from SOI, remains challenging.

It has been shown that the general DNG metamaterial exhibits distinct responses to TE and TM waves. Consequently, the interaction between the general DNG metamaterial and the nonparaxial incident CP components encompasses various SOI effects. Inherently, due to the differences in the dynamics of TE and TM fields upon reflection and transmission, the generated cross-polarized beam exhibits a vortex distribution. The physical interpretations and theoretical derivations of such a genuine SOI effect have been extensively discussed in [13]. Our method offers a powerful tool that provides more visualized information on the generated OV beam, facilitating a better under-standing of this intriguing phenomenon.

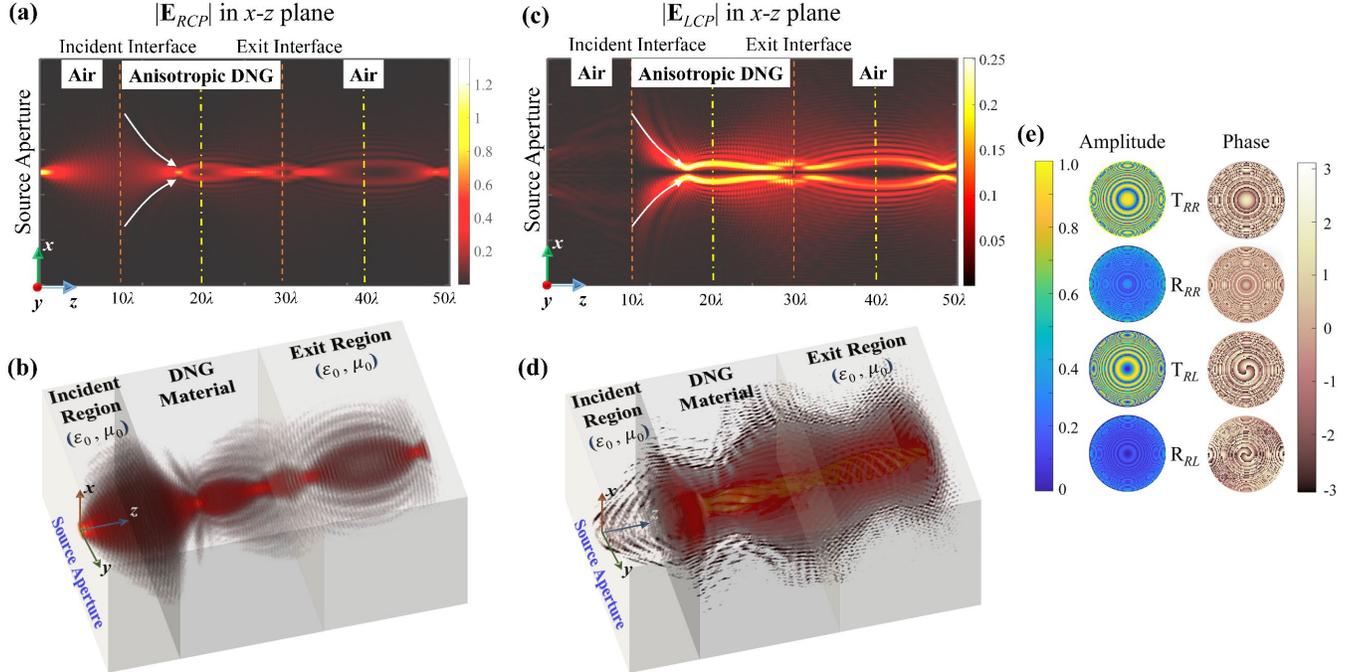

FIG.5. (a) The RCP wave's amplitude in the *x-z* plane across the whole system. (b) The 3D vision of the wavefront $|\mathcal{R}e(E_R)|$. (c) The coupled LCP wave's amplitude in the *x-z* plane across the whole system. (d) The 3D vision of coupled wavefront $|\mathcal{R}e(E_L)|$. (e) The amplitudes and phase of transmission and reflection coefficients for co-polarization and cross-polarization in momentum space.

The polarization of our source Gaussian beam is selected to be RCP, with its formula given as follows:

$$\mathbf{E}_{RCP}^{inci}(x,y,0) = \frac{1}{\sqrt{2}} e^{\frac{-(x^2+y^2)}{w_0^2}} [1 \quad i] \begin{bmatrix} \hat{\mathbf{e}}_\mathbf{x} \\ \hat{\mathbf{e}}_\mathbf{y} \end{bmatrix}. \quad (13)$$

It is worth re-emphasizing that the defined perfect CP beam in the transverse *x-y* plane appears elliptically CP when examined on the ***k***-normal planes, where the basis is TE and TM waves. This SOI effect is crucial for OV generation. The simulation results are depicted in FIG.5. To better observe the generated OV beam, the wavefront distributions across the whole system is illuminated in 3D perspective view.

As anticipated, both the co-polarized and cross-polarized beams exhibit dual focusing within the DNG metamaterial and free space, attributed to the responses of TE and TM waves. Notably, the majority of the cross-coupled power stems from momentum components with incident angles near the *Brewster* angle, as delineated by the white arrows in FIG.5(a) and (c). The fundamental mechanism is that the TE and TM waves being out of phase at the Brewster angle, facilitating efficient conversion. At last, the cross-coupled energy forms a ring-like band, as illustrated in FIG.5(e).

The OV beams carry orbital angular momentum (OAM), and therefore the normal momentum ($k_z = k_0$) is disallowed, resulting in a diminished "directivity" for OV beams. In some scenarios, the diffusivity of OV beams constrains their applicability, such as high-capacity multi-channel wireless communications. Figure 5 implies that the DNG metamaterial can effectively collimate the dispersing OV beam, thus presenting a potential candidate for enhanced OAM generation with superior diffusivity control.

Examination of the topological charge, *l*, of an OV beam is of significance. Studies [13-15] establish that *l* should follow the relation:

$$l = -2 \times q. \quad (14)$$

here *q* represents the polarization charge. For a $C_{4v}$ material, which corresponds to our provided DNG material, the polarization charge is 1. Consequently, the topological charge of the generated OV beam should be 2. To verify this assertion, we assess the reflection and transmission coefficients for the co-polarized and cross-polarized momentum components, as depicted in FIG.5(e). The phase distributions of cross-coupling coefficients display two cycles, confirming that the topological charge for both reflected and transmitted OV beams is 2.

Furthermore, the Supplemental Material [46] corroborates that the DNG near-zero-index material (NZIM) can effectively enhance vortex generation from a paraxial beam.

The Supplemental Multimedia Material [72] illustrates that the rotation direction of the wavefront of the vortex beam are opposite in the DNG material and the free space, which is a interesting phenomena that has never been reported before. Again, our proposed methodology enables the researchers to explore many unnoticed phenomena in a visualized way.

## IV. CONCLUSION

In this study, we introduce a novel methodology combining matrix Fourier optics and GTMM to elucidate intricate interactions between nonparaxial beams with various SOPs and bianisotropic metamaterials. We present the theoretical framework and demonstrate the method using an air-metamaterial-air system. By applying the proposed methodology, we investigate the interactions between a mismatched anisotropic DNG metamaterial and Gaussian beams with diverse SOPs. We report the bifocal phenomenon for linearly polarized incident beams and discuss the system in the laboratory coordinate system. Additionally, we analyze the generation of cross-coupled CP OV beams resulting from the SOI effect and offer visualizations of the complex fields. Most cross-coupled OV beam's energy is from the momentum components with incident angles near the *Brewster* angle is observed, and the topological charge of the generated OV beam is proven to be 2, aligning with other theoretical predictions.

Our methodology paves the way for exploring intriguing research avenues of metamaterials, such as circular birefringence and circular dichroism in chiral materials, and nonreciprocity in bi-anisotropic metamaterials, among others.

## ACKNOWLEDGEMENT

Y.F. Qin appreciates Dr. L. Kang, Dr. C. Peng and Dr. Y. J. Fu for fruitful discussions and valuable suggestions. Dr. L. Kang's careful review significantly improves the quality of this paper.